\documentclass[
reprint,
amsmath, amssymb,
aps,
]{revtex4-2}
\usepackage[version=3]{mhchem}
\usepackage{graphicx} 
\usepackage[usenames,dvipsnames]{xcolor}
\usepackage{hyperref}
\usepackage{physics}
\usepackage{comment}

\providecommand*{\pp}[3][]{\frac{\partial^{#1}#2}{\partial #3^{#1}}}

\providecommand*{\rmd}{\mathrm{d}}

\renewcommand{\phi}{\varphi}

\begin{document}

\title{Non-monotonic frictional behavior in the lubricated sliding of soft patterned surfaces} 

\author{Arash Kargar-Estahbanati} 
\author{Bhargav Rallabandi}
\email{bhargav@engr.ucr.edu}
\affiliation{Department of Mechanical Engineering,
 University of California, Riverside, CA 92521
}

\begin{abstract} 
    We study the lubricated contact of sliding soft surfaces that are locally patterned but globally cylindrical, held together under an external normal force. The local patterns represent either naturally occurring surface roughness or engineered surface textures. Three dimensionless parameters govern the system: a speed, and the amplitude and wavelength of the pattern. Using numerical solutions of the Reynolds lubrication equation, we investigate the effects of these dimensionless parameters on key variables such as contact pressure and the coefficient of friction of the lubricated system. For small roughness amplitudes, the coefficient of friction increases with roughness. However, our findings reveal that increasing surface roughness beyond a critical value can decrease the friction coefficient, a result that contradicts conventional intuition and classical studies on the lubrication of rigid surfaces. For very large roughness amplitudes, we show that the coefficient of friction drops even below the corresponding smooth case. We support these observations with a combination of perturbation theory and physical arguments, identifying scaling laws for large and small speeds, and for large and small roughness amplitudes. This study provides a quantitative understanding of friction in the contact of soft, wet objects and lays  theoretical foundations for incorporating the friction coefficient into haptic feedback systems in soft robotics and haptic engineering. 
\end{abstract}

\maketitle

%%%END OF FOOTNOTES%%%

%%%MAIN TEXT%%%%
\section{Introduction}
A lubricant, a viscous liquid inserted between two moving objects in close contact, plays a vital role in minimizing friction and extending the lifespan of engineering components \citep{Okrent_1961,borsoff1959,mcgeehan1978}. 
The growing number of applications of soft materials in recent years has drawn renewed interest in ``soft lubrication''. For fluid-separated soft objects in close proximity to each other, the pressure due to flow is sufficiently strong to significantly deform at least one of the contacting surfaces \citep{skotheim2004,saintyves2016}. This lets the lubricated film support external forces and torques \cite{rallabandi2024ARFM}. This phenomenon holds significance in various areas, including the study of soft robotic hands \citep{peng2021}, prosthetic synovial joints \citep{hou1992}, and the interaction between eyeballs and contact lenses \citep{jin2005}. A characteristic feature of these systems is that the thickness of the separating fluid film is set dynamically by the relative velocity between the surfaces (faster motion typically leads to thicker films).

Soft lubricated systems typically reside in one of three distinct regimes: the ``boundary regime'', where the lubricant is too thin to support any load, and most of the load is borne by localized dry contacts; the ``mixed regime'', where the lubricant partially supports the load, but some portion is still carried by dry solid-to-solid contact; and the ``elasto-hydrodynamic lubrication'' (EHL) regime, where a robust film completely separates the contacting surfaces. This study  focuses exclusively on the EHL regime, characterized by the absence of solid-solid contact between lubricated surfaces \citep{Lu_2006, jacobson2003}.

The strong coupling between the elasticity of the contacting bodies and the flow of the lubricating fluid enables EHL systems to withstand both normal and tangential forces. However, this connection adds complexity to the theoretical analysis of such systems, often requiring numerical methods and analytical analysis is only feasible under asymptotic conditions \citep{bissett1989I, bissett1989II}.  
In the case of small normal loads, referred to as "the non-conformal regime", the surfaces move with a relatively thick fluid film, experiencing small deformation  \cite{sekimoto1993,skotheim2004}. Conversely, for large normal forces, the geometry resembles classical Hertzian contact, separated by a thin fluid film in what is termed the "conformal contact regime"\cite{essink2021,hui2021}. There may be a transition between the two regimes depending on the flow speed. Additionally, it is common to use different elastic formulations for the soft material  depending on the thickness of the soft substrate relative to the flow length scale. The widely-used Winkler model is best  suited to thin, compressible soft coatings \citep{skotheim2004, essink2021}, while for very thick materials, it is typical to treat the soft material as an elastic half-space \citep{skotheim2005,snoeijer2013}. More sophisticated analyses for intermediate thickness bridge the gap between these limits \citep{kargar2021,chandler2020}. In experimental studies, atomic force microscopy and a non-contacting probe have been used to measure EHL forces \citep{leroy2012,zhang2020}, while optical interferometry have characterized fluid film thickness and soft surface deformation profiles \citep{saintyves2016,davies2018}.

Many lubrication studies have focused on the measurement of effective friction force between lubricated soft surfaces. Previous studies have observed that altering surface architecture, for example,  by adding geometric micropatterns \citep{mourier2006,touche2016,suh2010} or periodically modifying stiffness \citep{moyle2020,moyle2022,wu2022}, profoundly influences tribological performance of  soft lubricated contacts. Additionally, soft surfaces in nature often exhibit roughness at the nanoscale \citep{ayyildiz2018}, underscoring the importance of understanding frictional behavior of lubricated uneven surfaces in fields like soft robotics \citep{adams2013}, and haptic engineering \citep{basdogan2020}. Experimental studies have found that surface asperities affect the shape of the lubricating film \cite{kaneta1980effects, choo2008interaction} and affect force transmission \cite{smith2002, peng2021}. \citet{peng2021}, in  particular, show that the friction between soft patterned surfaces appears to depend non-monotonically on the sliding speed. Wu et al. (2022) \citep{wu2022} undertook a numerical study for surfaces with periodically varying stiffness, correlating the additional friction force with increased energy dissipation compared to smooth surfaces. However, detailed theoretical and numerical studies of the effect of  geometrical patterns of the EHL films and force transmission is lacking. 
%Building upon experimental observations and scaling analysis by Peng et al. (2021) \citep{peng2021}, our study introduces a theoretical framework to EHL systems with geometric patterns, a first of its kind.  

In this study, we investigate the elastohydrodynamic lubrication of surfaces with geometric patterns by developing numerical solutions to the coupled fluid-elastic problem. Focusing on sinusoidal patterns on a two-dimensional cylindrical contact geometry, we study the effect of sliding speed, the wavelength and amplitude of the pattern. Of particular interest is the effective coefficient of friction (the ratio between the EHL drag and the applied normal load). Our findings show that for small roughness the friction increases with the square of the roughness, while for very rough surfaces, the friction coefficient decreases as roughness increases. We explore the physical basis of this non-trivial observation using scaling analysis and perturbation theory. 

\section{Problem formulation} \label{SecFormulation}

\subsection{Setup and governing equations}
\begin{figure}
    \centering
    \includegraphics[width=0.4 \textwidth]{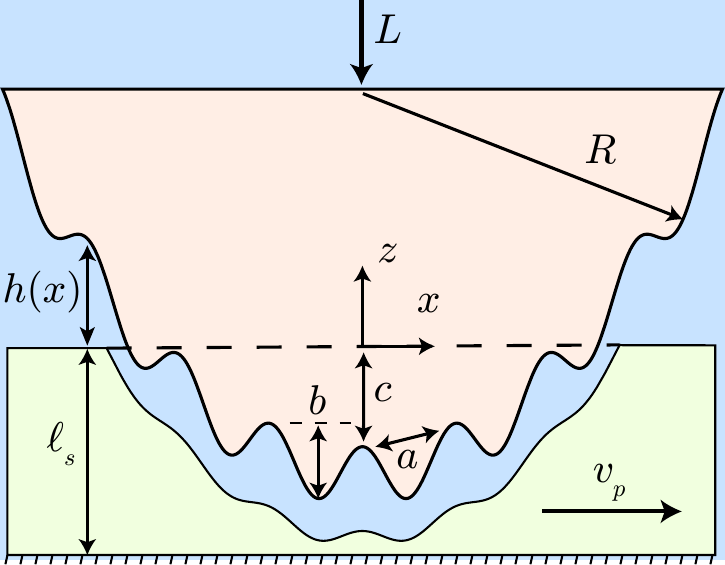}
    \caption{ Sketch of the geometry and coordinates; a patterned surface with asperities is submerged in a fluid and pressed into a deformable soft coating, which is moving to the right.  The dashed line represents the original, undeformed position of the soft coating.}
    \label{Schematic}
\end{figure}

We consider the sliding motion of a patterned cylinder pressed into a plane coated with an elastic material of thickness $\ell_s$ under a normal load $L$ (per length of the cylinder). The two surfaces are submerged in a Newtonian fluid with viscosity $\eta$, and the bottom surface translates from left to right with velocity $v_p$, as depicted in Fig. \ref{Schematic}. We assume the coating to be linear elastic with shear modulus $G$ and Poisson's ratio $\nu$. Relative motion establishes a layer of fluid with thickness $h(x)$ between the  bodies, forming a ``lubricated contact''. The cylinder radius $R$ is considerably larger than the fluid film thickness $h(x)$. We also focus our attention to sinusoidal patterns of amplitude $b/2$ and half-wavelength $a$, though more general patterns have been studied in experiments and modeling efforts \citep{ayyildiz2018}. %Thus, the patterned cylinder surface is locally approximated as a sinusoidally-perturbed parabola $x^2/(2R) + \frac{b}{2} \cos{\pi x/a}$. 

The flow is assumed to be incompressible and at steady state and the effect of inertia of the flow is assumed negligible. We further consider gently varying patterns such that $b \ll a$. Thus, we describe the flow using the thin-film equation, and define velocity in the thin film as
\begin{equation}\label{velocity_eq}
    v_x=\frac{1}{2\eta} \frac{\partial p}{\partial x} (z+\delta)(z+\delta-h)+ v_p \left(1- \frac{z+\delta}{h}\right)
\end{equation}
where $p(x)$ is the fluid pressure and $\delta(x)$ is deformation of the elastic coating (positive when the coating is depressed). Integrating \eqref{velocity_eq} across the fluid gap, and demanding that the flux is constant at steady state leads to Reynolds lubrication equation for the pressure
\begin{equation}\label{lubrication_eq}
    \pdv{}{x} \qty( h^3 \pdv{p}{x} - 6 \eta v_p h )= 0 .
\end{equation}
The film thickness $h(x)$ is a function of deformation $\delta(x)$ and geometry of the pattered cylindrical surface. %top surface $x^2/(2R) + (b/2) \cos{\pi x/a}$. 
As the surfaces are pressed together, the nominal cylindrical surface penetrates the coating a distance $c < \ell_s$. Using the standard parabolic approximation of the cylinder, valid around the central axis $(x = 0)$, the top surface is at a location $z = h_{\rm top}(x) = x^2/(2R) + (b/2)\cos{(\pi x/a)} -c$, while the surface of the deformed elastic coating is at $z = -\delta(x)$. Thus, the film thickness is 
\begin{equation}\label{Geometry}
h(x) = \frac{x^2}{2R} + \frac{b}{2} \cos{\frac{\pi x}{a}} - c + \delta(x).
\end{equation}
 We note that the ``penetration depth'' $c$ depends on the sliding speed and is determined dynamically as part of the solution. It is positive for static ``dry'' contact and small sliding speeds (as shown in Fig. \ref{Schematic}) and may become negative at large speeds, corresponding to a cylinder that sits ``raised above'' the coating with a clearance between the undeformed surfaces. %Note that by definition $c$ is the penetration depth at $x=0$, constant for a given velocity $v_p$. 

The deformation $\delta(x)$ is related to the fluid pressure $p(x)$ \textit{via} the elastic response of soft substrate. We focus here on thin compressible elastic coatings where $\ell_s \ll \sqrt{R |c|}$. Then,  the surface deformation at any point on the coating is locally related to the fluid pressure at the same point \citep{johnson1987} 
\begin{equation}\label{elastic_response}
    \delta(x) = K p(x), \quad \mbox{where}\quad  K = \frac{\ell_s (1-2\nu)}{2G (1-\nu) }\,
\end{equation}
is the Winkler elastic compliance. 

Subject to vanishing condition for pressure far outside the film ($p(\pm \infty) =0$), equations \eqref{lubrication_eq} and \eqref{Geometry} yield a system of equations for $p(x)$ and $h(x)$. In addition, the penetration depth $c$ must be determined self consistently such that the stress of the flow counteracts the applied normal load (per length) $L$, 
\begin{equation}\label{lift}
    L=\int_{-\infty}^{\infty} p\,dx\,.
\end{equation}
The system of equations \eqref{lubrication_eq}--\eqref{lift}, once solved simultaneously, yield the pressure $p(x)$, film thickness $h(x)$ and penetration depth $c$ in terms of $v_p$, $L$ and the mechanical and geometrical properties of the system.

In the static limit with $v_p=0$, the fluid drains from the gap between the contacting surfaces, resulting in $h(x)=0$. Thus, equations \eqref{Geometry} becomes independent of the lubrication equation \eqref{lubrication_eq}. This scenario, known as "dry contact'', requires solving \eqref{Geometry}, \eqref{elastic_response}, and \eqref{lift} simultaneously to determine pressure $p(x)$ and penetration depth $c$ (see \citep{johnson1987} for details). We later use the dry contact problem to evaluate the quasi-static limit, where the relative velocity is small, and the fluid film is extremely thin.

%%%%%%
%%%%%%%%%%%%%
\subsection{Non-dimensionalization}
We rescale the governing equations before solving the problem, using the ``dry'' static contact problem in the smooth limit $(b = 0)$ to identify characteristic length and pressure scales. This limit admits an analytic solution.  
The two surfaces make contact in the region $-\ell < x < \ell$, where  
\begin{equation} \label{smooth_and_dry_ell}
\ell = \left( \frac{3LR\ell_s}{4G} \frac{1 - 2\nu}{1 - \nu} \right)^{1/3}
\end{equation}
is the static contact half-length. The contact pressure distribution in this limit is 
\begin{equation}  \label{smooth_and_dry_p}
    p(x) =  \frac{3 L}{4 \ell} \left(1 - \frac{x^2}{\ell^2}\right) \quad \mbox{(smooth, dry)}. 
\end{equation}
Thus, in the dry, smooth limit, the total contact length is $2\ell$, the maximum pressure is $\frac{3L}{4\ell}$, and the vertical length scale (which sets the magnitude of $c$) is $\ell^2/2R$. We use these scales to define  dimensionless quantities (denoted with overbars) as follows: 
\begin{equation}\label{rescaled_vars}
    \overline{x}=\frac{x}{\ell}, \qquad  \overline{h}=\frac{h}{\ell^2/(2R)}, \qquad  \overline{c}=\frac{c}{\ell^2/(2R)}, \qquad  \overline{p}=\frac{p}{3L/4\ell}\,.
\end{equation}
Substituting these scales into the Reynolds equation identifies a dimensionless velocity 
\begin{align}
    \lambda = \frac{32 \eta v_p R^2}{L \ell^2}  =  \frac{2^{19/3}}{3^{2/3}} \frac{\eta v_p R^{4/3}}{L^{5/3} \ell_s^{2/3}} \left(\frac{G(1-\nu)}{1-2 \nu}\right)^{2/3}.
\end{align} 
The system of governing equations \eqref{lubrication_eq}--\eqref{lift} rescale as
\begin{subequations}\label{ODE_system}
    \begin{equation}\label{lubrication_norm}
        %\frac{\partial \overline{p}}{\partial \overline{x}}= \lambda \frac{\overline{h}- \overline{h}^*}{\overline{h}^3},
        \pdv{}{\overline{x}} \qty(\overline{h}^3 \pdv{\overline{p}}{\overline{x}} - \lambda \overline{h})=0,
    \end{equation}
    \begin{equation}\label{Geom_norm}
        \overline{h}(\overline{x})= \overline{p}+ \overline{x}^2+ \frac{\beta}{2} \cos{\frac{\pi \overline{x}}{\alpha}} - \overline{c},
    \end{equation} 
    \begin{equation}\label{lift_norm}
        \int_{-\infty}^{\infty} \overline{p}\,d\overline{x}=\frac{4}{3},
    \end{equation}
\end{subequations}
where $\beta=2Rb/\ell^2$ and $\alpha=a/\ell$ are rescaled roughness half-amplitude and half-wavelength, respectively. In this study, we focus on roughness amplitudes comparable to the penetration depth ($\beta \sim 1$) and wavelengths that are significantly shorter than the contact length ($\alpha \ll 1$). 

We solve the system of equations \eqref{ODE_system} by initially guessing a value for penetration depth $\overline{c}$. The equations \eqref{lubrication_norm} and \eqref{Geom_norm}, which yield a second-order nonlinear ODE subjected to $\overline{p}(\pm \infty)=0$, are solved using the shooting method so that the pressure decays to zero at both ends of the computational domain. Due to the multiple scales of the problem, precise resolution is needed in regions of high curvature, while other regions can be solved with a coarser grid. We therefore employed MATLAB's ODE45 solver, which automatically generates a non-uniform grid based on error tolerance and solution smoothness. The next section demonstrates that the solution profile supports this approach.  At this stage the solution only satisfies the flow equation \eqref{lubrication_norm}, while the normal load balance \eqref{lift_norm} is not yet met. Therefore, we iterate on $\overline{c}$ (re-solving the lubrication problem each time) to satisfy \eqref{lift_norm} within a small tolerance, completing a solution to the system \eqref{ODE_system} for a prescribed set of $\lambda$, $\alpha$ and  $\beta$. 
%We used a finite difference method on a non-uniform grid to solve this system of equations.

%%%%%%fig 1
\begin{comment}
 \begin{figure}[h!]
\begin{subfigure}{0.45\textwidth}
     \centering
      \includegraphics[width=0.9\linewidth]{FilmThickness_beta=1.pdf}
      \label{fig.filmthickness}
      %\caption{}
\end{subfigure}
\begin{subfigure}{0.45\textwidth}
     \centering
      \includegraphics[width=0.9\linewidth]{Pressure_beta=1.pdf}
      \label{fig.pressure}
      %\caption{}
\end{subfigure}
\caption{(a) Dimensionless film thickness ($\overline{h}$) and (b) dimensionless pressure ($\overline{p}$) plotted against the dimensionless horizontal coordinate ($\overline{x}$) for various values of normalized speed $\lambda=5,0.5,\textrm{and},0.05$. Throughout the figure, $\beta=1$ and $\alpha=0.1$.}
\label{fig.h and p}
\end{figure}   
\end{comment}

\begin{figure}[t!]
    \centering
    \includegraphics[width=0.4\textwidth]{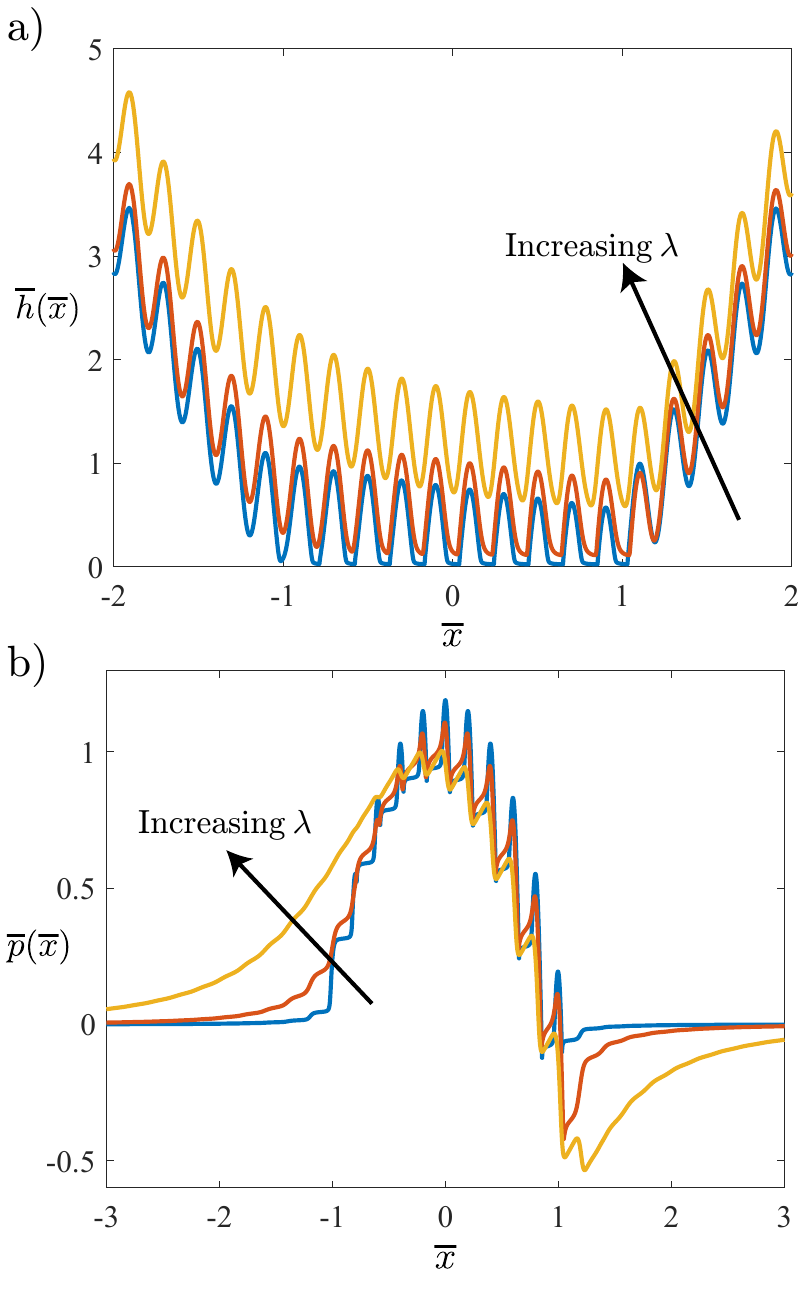}
    \caption{(a) Dimensionless film thickness ($\overline{h}$) and (b) dimensionless pressure ($\overline{p}$) plotted against the dimensionless horizontal coordinate ($\overline{x}$) for various values of normalized speed $\lambda=5,0.5,\textrm{and},0.05$. Throughout the figure, $\beta=1$ and $\alpha=0.1$.}
    \label{fig.h and p}
\end{figure}
%%%%%%%%%%%%%%%%%%%%%%
\section{Results and Discussion}
%In this section, we expand on the elastohydrodynamic problem introduced in the formulation. First, 
%\sout{We examine the solutions of the system of equations \eqref{ODE_system}, the rescaled pressure ($p/(3L/4\ell)$), and the rescaled film thickness ($h/(\ell^2/2R)$) across a range of dimensionless parameters $\lambda$, $\alpha$, and $\beta$.} \BR{[[I moved this to the next section]]} We then analyze the impact of these parameters on the rescaled friction coefficient $(\mu\,2R/\ell)$. Additionally, we investigate counter-intuitive behaviors observed under extreme conditions and provide analytical insights using perturbation theory. \BR{Do we need this paragraph? I think we can just go ahead and discuss everything since at this point the reader doesn't have sufficient intuition or context for them to appreciate ``counter-intuitive behaviors observed under extreme conditions''. }

%%%%%%%%%%%%%%%
\subsection{Pressure and film thickness}
We examine solutions of the system of equations \eqref{ODE_system}, the rescaled pressure ($p/(3L/4\ell)$), and the rescaled film thickness ($h/(\ell^2/2R)$) across a range of dimensionless parameters $\lambda$, $\alpha$, and $\beta$. Figure \ref{fig.h and p} plots the fluid film thickness ($\overline{h}$) and  pressure ($\overline{p}$) as  functions of the  horizontal coordinate ($\overline{x}$) for different values of normalized speed ($\lambda$). The surface roughness $\beta$ and the rescaled amplitude of asperities $\alpha$ are constant throughout Fig. \ref{fig.h and p}. The variables $\overline{p}$ and $\overline{h}$ are directly derived from the numerical solution of \eqref{ODE_system}, and the precision of their calculation determines the accuracy of our results. Thus, a mesh study was performed to assure that the final solution is converged. 
Fig. \ref{fig.h and p} confirms that the problem has multiple scales, necessitating a non-uniform mesh to accurately resolve the profile. 
For small $\lambda$, representing a small speed or a large normal load, the fluid film thickness and pressure distribution are strongly influenced by the roughness.  In particular, we see localized regions with larger pressure and very thin films (small $\overline{h}$), which correspond to the regions of contact inherited from the static problem. These are separated by ``gaps'' where the fluid film is much thicker. %These features, as we will later show, are important in understanding the dependence of the friction on the % One can also observe that, as a result of large normal force and high fluid pressure, the film thickness is determined by deformation of the soft substrate ($\delta$), which results in some non-trivial features in the behavior of the friction coefficient,  discussed later in this section.
For large speeds ($\lambda \gg 1$) by contrast, $\overline{h}$ significantly exceeds $\beta$. Consequently, the solution to the elasto-hydrodynamic problem predominantly mirrors the smooth-cylinder solution, augmented by relatively minor fluctuations resulting from the sinusoidal protrusions. %Later in this section, we use a perturbative expansion to develop analytical results for the friction coefficient as a function of the rescaled parameters in this limit. 

%%%%%%%%%%%%%%%
\subsection{Coefficient of friction}
%%%%%%%%%%%%%
%%%%%%%% fig 2 %%%%%%%%
\begin{comment}
 \begin{figure}[t!]
\begin{subfigure}{0.45\textwidth}
     \centering
      \includegraphics[width=0.90\linewidth]{fig_mu_lmbd_alpha_2.pdf}
      \label{fig_mu_lmbd_alpha}
      %\caption{}
\end{subfigure}
\begin{subfigure}{0.45\textwidth}
     \centering
      \includegraphics[width=0.9\linewidth]{mu_lambda_more_points.pdf}
      \label{fig_mu_lambda}
      %\caption{}
\end{subfigure}
\caption{ (a) Friction coefficient ($\mu$) versus dimensionless speed ($\lambda$) for different values of dimensionless wave length ($\alpha$), $\beta = 0.2$  (b) Friction coefficient ($\mu$) versus dimensionless speed ($\lambda$) for different values of roughness ($\beta$), $\alpha=0.1$.}
\label{fig.mu_lambda_alpha}
\end{figure}   
\end{comment}

\begin{figure}[t!]
    \centering
    \includegraphics[width=0.4\textwidth]{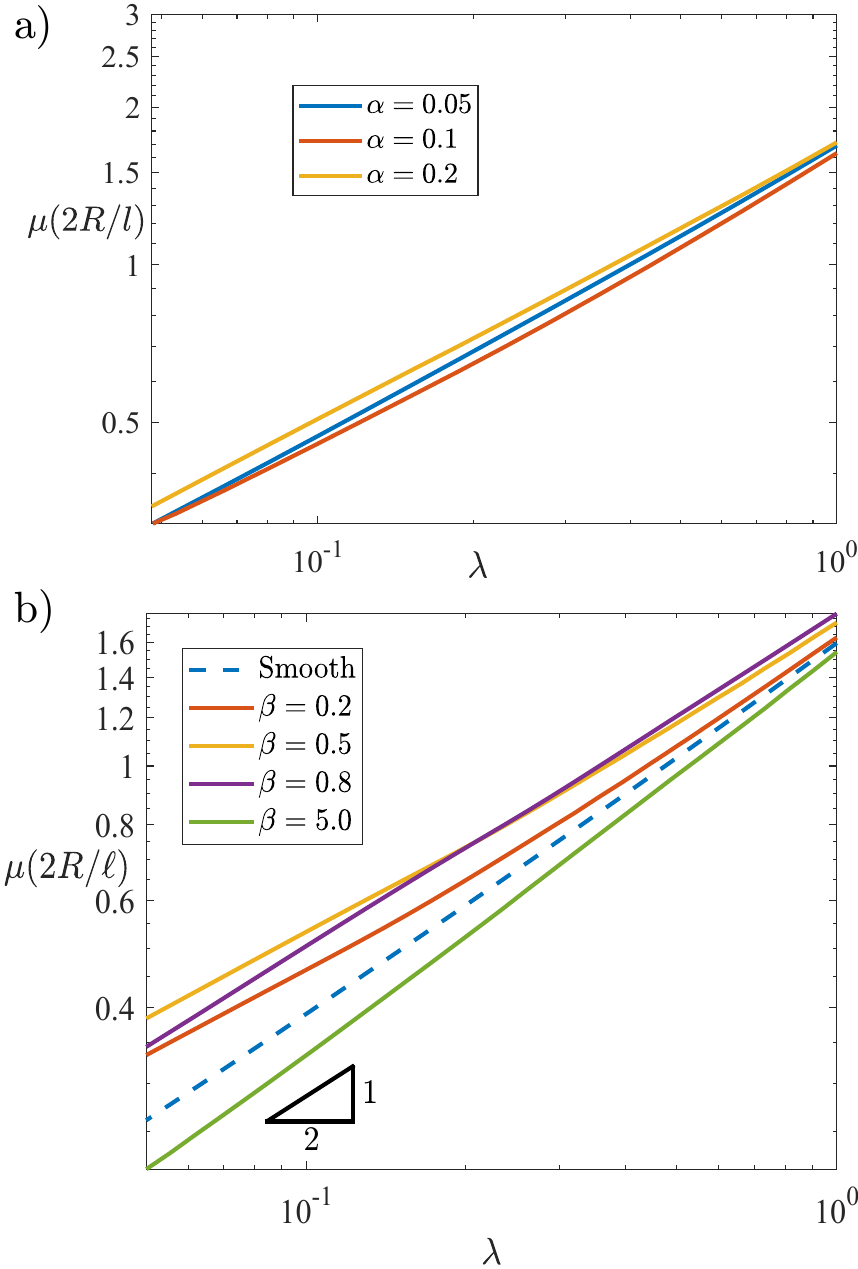}
    \caption{(a) Friction coefficient ($\mu$) versus dimensionless speed ($\lambda$) for different values of dimensionless wave length ($\alpha$), $\beta = 0.2$  (b) Friction coefficient ($\mu$) versus dimensionless speed ($\lambda$) for different values of roughness ($\beta$), $\alpha=0.1$.}
    \label{fig.mu_lambda_alpha}
\end{figure}

%%%%%%%%%%%%
As noted earlier, the primary objective of this study is to understand the effect of surface roughness on the coefficient of friction $\mu$, defined as the ratio of the EHL drag $D$ to the applied load $L$. From solutions of the pressure and film thickness, $\mu$ is found as %the ratio between the tangential force (drag, $D$) and the applied load $L$, 
\begin{subequations} \label{friction}
    \begin{align}
        \mu= \frac{D}{L} &= \frac{1}{L} \int \left(-p \pp{h}{x} - \eta \pp{v}{z} \bigg|_{z = h_{\rm top}}\right) dx, \\
        & = \frac{\ell}{2R}\frac{\displaystyle\int\frac{1}{2}\overline{h}\frac{\partial \overline{p}}{\partial \overline{x}} -\frac{\lambda}{6 \overline{h}}+ \frac{\partial \overline{h}_{\rm top}}{\partial \overline{x}} \overline{p}\, d\overline{x}}  {4/3},
    \end{align}
\end{subequations}
where (\ref{friction}b) uses rescaled variables. We note that since the load is held constant throughout in the present formulation (independent of the parameters $\alpha$, $\beta$ and $\lambda$), the coefficient of friction is simply a rescaled version of the drag. 

We first study the effect of the protrusion amplitude and wavelength on the friction coefficient. The wavelength of the protrusions $2\alpha$ evidently affects the spatial structure of the pressure and film thickness distributions (Fig. \ref{fig.h and p}). However, we find that the wavelength has only a weak effect on the friction coefficient, particularly when it is much smaller than the contact length ($\alpha \ll 1$), as illustrated in Fig. \ref{fig.mu_lambda_alpha}a. A similar independence of forces on the wavelength of surface patterns was shown to hold (using formal asymptotic arguments and validated numerically) in the context of rigid surfaces with sinusoidal undulations \citep{Yariv2024}. Later, we offer a rationalization of this independence using scaling and symmetry arguments, accounting for deformations. The condition of small wavelength is natural to consider in the context of ``surface textures'' or ``surface roughness'', whereas the case $\alpha = O(1)$ modifies the cylindrical shape across the entire lubricated film region and is not the regime of interest here. We thus work in the small-wavelength regime throughout the paper, focusing on the dependence of $\mu$ on the remaining dimensionless parameters, $\lambda$ and $\beta$. 

Figure~\ref{fig.mu_lambda_alpha}b studies the influence of dimensionless speed $\lambda$ on friction coefficient for different values of the rescaled surface roughness $\beta$. For a smooth cylinder at small $\lambda$, a balance between the $O(1)$ pressure gradient, determined by dry contact, and the viscous stress in \eqref{lubrication_norm}, suggests a film thickness scaling of $\overline{h}_0^* \propto \lambda^{1/2}$ (see also  Ref. \cite{kargar2022}). By substituting the film thickness and pressure scaling into \eqref{friction}b, the friction coefficient is found to scale as $\mu \propto \ell / 2R \lambda^{1/2}$. This scaling with $\lambda$ is consistent with the numerical results for $\beta = 0$, shown in Figure~\ref{fig.mu_lambda_alpha}b.
As the roughness amplitude $\beta$ increases from zero at a fixed speed $\lambda$, the friction coefficient increases, aligning with both experimental observations \cite{peng2021} and intuitive expectations. However, the opposite is observed for sufficiently large $\beta$ -- the coefficient of friction \emph{decreases} with increasing roughness. This non-monotonic dependence of the friction on the roughness amplitude is most noticeable for small and moderate speeds $\lambda$, where the deformation of the solid is comparable to (or greater than) the film thickness; compare the curves for $\beta = 0.8$ and $\beta = 0.5$ in Fig. \ref{fig.mu_lambda_alpha}b. Surprisingly, for very large $\beta$, the coefficient of friction can drop to values even below the smooth limit ($\beta = 5$ curve in Fig. \ref{fig.mu_lambda_alpha}b). 

%%%%%%% fig 3 %%%%%%%%
\begin{figure}
     \centering
      \includegraphics[width=0.95\linewidth]{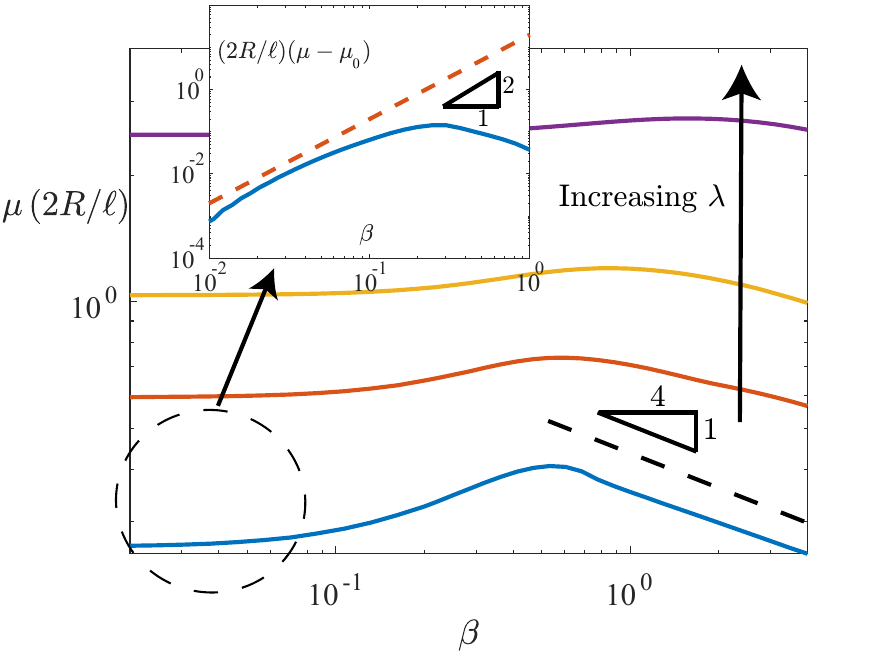}
      \label{fig.mu_beta}
      %\caption{}
%\begin{subfigure}{0.45\textwidth}
%     \centering
%      \includegraphics[width=0.95\linewidth]{FilmThickness_Zoomed_Edit_V2.pdf}
%      \label{fig.h zoomed}
      %\caption{}
%\end{subfigure}
\caption{ Friction coefficient versus roughness ($\beta$) for different values of dimensionless speed $\lambda=0.05,\, 0.2,\, 0.5,\, 2.0$. The figure inset shows a plot of the difference between the friction coefficients in the rough and smooth cases ($\mu - \mu_{0}$) for small roughness amplitude $\beta \ll 1$. The dashed line represents the scaling analysis, while the solid line shows the numerical results.}
\label{fig. mu_lambda}
\end{figure}
%%%%%%%%%%%%%
%%%%%

To illustrate this counter-intuitive behavior more clearly, in Fig. \ref{fig. mu_lambda} we plot the rescaled friction coefficient $\mu (2R/\ell)$ versus roughness $\beta$. Starting from a smooth surface ($\beta=0$), increasing roughness $\beta$ first results in an increase in the friction coefficient. A maximum occurs at critical value of $\beta$, beyond which increasing the roughness \emph{lowers} the friction coefficient. Eventually, $\mu$ falls below its smooth limit. %Thus, for a fixed $\lambda$, the coefficient of friction is maximized at a critical $\beta$. 
As $\lambda$ increases, the value of this critical $\beta$ (corresponding to the maximum $\mu$ for that $\lambda$) increases slightly, while the height of the maximum itself diminishes. These features seem to persist for larger $\lambda$ but become less prominent.

\subsection{Friction at low speeds: scaling analysis} \label{SecSmallLam}
{We rationalize these observations using symmetry and scaling arguments, focusing separately the limits of large and small values of either $\beta$ or $\lambda$. We first consider the limit of small $\lambda$, which displays the strongest dependence of $\mu$ on $\beta$. For small $\beta$, it is useful to perturb around the smooth solution, where the film thickness is $\overline{h}_{0}(\overline{x})$ and has a characteristic scale $\overline{h}_0^*$. For $\beta \ll \overline{h}_0^*$, we expect that the film thickness takes the shape $\overline{h}(\overline{x}) = \overline{h}_{0}(\overline{x}) (1 + \sigma \mathcal{S} + \sigma^2 \mathcal{S}^2 +  O(\sigma^3))$, where $\sigma = \beta/\overline{h}_0^*$ is a small parameter and $\mathcal{S}(\overline{x})$ is a sinusoidal function of $O(1)$ amplitude and $O(\alpha) \ll 1$ wavelength. We now consider the contribution to the shear stress to the drag, which scales as $\int \lambda/ \overline{h} d\overline{x}$. We expand the integrand in powers of $\sigma$ as $\int \lambda/ \overline{h}_0 (1 + \sigma \mathcal{S} + \sigma^2 \mathcal{S}^2) d\overline{x}$ (note that we omit prefactors in front of the powers of $\sigma$ since the argument is at the level of scaling). The terms linear in $\epsilon$ are also linear in the rapidly oscillating sinusoid $\mathcal{S}(\overline{x})$, so it integrates to zero. The first nonzero roughness contribution results from the term involving $O(\sigma^2 \mathcal{S}^2)$, which on integration yields a term of $O(\sigma^2)$ that is \emph{independent} of $\alpha$ (this independence occurs because the integral picks out the zero-wavenumber Fourier component of $\mathcal{S}^2$). A similar symmetry argument also applies to terms involving the pressure. Thus, we expect that the coefficient of friction for small $\beta$ takes the form
\begin{align} \label{SmallBetaAsy}
    \mu \sim \mu_{0}\left(1  + k \left(\frac{\beta}{\overline{h}_0^*}\right)^2\right) \quad \mbox{for} \quad \beta \ll \overline{h}_0^*, 
\end{align}
where $\mu_0$ is the smooth $(\beta = 0)$ limit, and the prefactor $k$ results from the detailed integration.  As noted earlier in the text, $\overline{h_0^*} \propto \lambda^{1/2}$ and $\mu_{0} \propto \ell / 2R \lambda^{1/2}$. Substituting these expressions into \eqref{SmallBetaAsy}, we expect that the friction coefficient for small $\beta$ and small $\lambda$ behaves as
\begin{align} \label{SmallLamBeta}
    \mu \sim \frac{\ell}{2R} \left( k_1 \lambda^{1/2} + k_2 \beta^2 \lambda^{-1/2} \right) \quad \mbox{for} \quad \beta \ll \lambda^{1/2} \ll 1, 
\end{align}
for $O(1)$ constants $k_1$ and $k_2$. The first term represents the smooth limit, while the second is the leading effect of the roughness. The difference between the rough and smooth friction coefficients scales as $\beta^2$ for small $\beta$, in agreement with numerical calculations (inset of Fig. \ref{fig. mu_lambda}). 

%\BR{[[In some places it is not so obvious what is dimensional and what is dimensionless. Please make consistent. I vote we use $\overline{\ell}_f$ for the dimensionless thin film contact length.]]}
We now analyze the scaling behavior for large roughness $(\beta \gg 1)$, still considering small speeds $(\lambda \ll 1)$. For large $\beta$, the dry contact problem is significantly different from the smooth case. The contact between the two surfaces is no longer contiguous, and is instead made up of $n$ isolated contact regions, where the individual protrusions locally deform the soft substrate. These contact regions are separated by gaps between the protrusions (see Fig. \ref{fig.dry}b). The contact regions are relatively narrow and have widths $\overline{\ell_f} \ll \alpha$, while the gaps have widths of $\ell_g = O(\alpha)$. When motion is initiated, the contacting regions become lubricated and support thin fluid films, while the gap regions remain largely unchanged, as illustrated schematically in Figure \ref{fig.dry}a. To understand the contributions of the two kinds of regions on the force, we define the cumulative drag and lift functions, respectively,
\begin{align}
    \mathcal{D}(\overline{x}) &=\int_{-\infty}^{\overline{x}} \left(\frac{1}{2}\overline{h}\frac{\partial \overline{p}}{\partial \overline{x}} -\frac{\lambda}{6 \overline{h}}+ \frac{\partial \overline{h}_{\rm top}}{\partial \overline{x}}\right) ds \\
    \mathcal{L}(\overline{x}) &=\int_{-\infty}^{\overline{x}} \overline{p}\, \rmd s,
\end{align}
where the integrands are functions of the ``dummy'' spatial variable $s$. Note that by definition $\mathcal{D}(\infty) = D$ and $\mathcal{L}(\infty) = 4/3$. 

Figure \ref{fig.cum_int} shows the accumulation of drag and lift forces in both the film and gap regions for $\lambda = 0.05$, $\beta = 4$. It is clear that the drag is accumulated in the $n$ thin films of length $\overline{\ell_f}$ corresponding to initially dry contact. These are regions of high shear stress (due to thin films) whereas the gaps have much thicker films and thus much lower stress. In contrast to the drag, the lift is accumulated gradually across the entire ``contact region'' of dimensionless length $\overline{\ell} \sim 1$, encompassing both gap and film regions. Since the films are much narrower than the gaps, most of the lift is therefore supported by the pressure in the gaps. Using \eqref{lift} to balance normal stresses with the applied load suggests a pressure scale of $O(L/\ell)$.  In dimensionless terms, we thus expect $\overline{p} = O(1)$ and $d \overline{p}/d \overline{x} = O(1)$ for the lubricated problem. We verify this conclusion by plotting the pressure for $\beta = 4$ and $\lambda = 0.05$ in Figure \ref{fig.dry}b. Contrary to the corresponding dry contact where the pressure is concentrated in the contact regions, the pressure in the lubricated patterned case remains relatively close to the smooth problem, with smaller fluctuations in the thin-film regions. We posit that this is due to the continuous fluid film, which distributes the pressure uniformly through requirements of continuity and a constant fluid flux through the system. 
\begin{comment}
\begin{figure}[t!]
\begin{subfigure}{0.4 \textwidth}
    \centering
    \includegraphics[width=\textwidth]{FilmThickness_Zoomed_Gap_Film_Schematic.pdf}
\end{subfigure}
\begin{subfigure}{0.4 \textwidth}
    \centering
    \includegraphics[width=\textwidth]{dry_smooth_wet.pdf}
\end{subfigure}
    \caption{ (a) A zoomed-in plot of the film thickness for $\beta=1.6$ and $\lambda=0.05$, highlighting the film and gap regions. (b)dimensionless pressure $\overline{p}$ versus dimensionless horizontal coordinate $\overline{x}$ for the $\beta = 4$ and $\lambda = 0.05$, the corresponding dry contact, and the dry smooth contact.}
    \label{fig.dry}
\end{figure}     
\end{comment}

\begin{figure}[t!]
    \centering
    \includegraphics[width=0.45\textwidth]{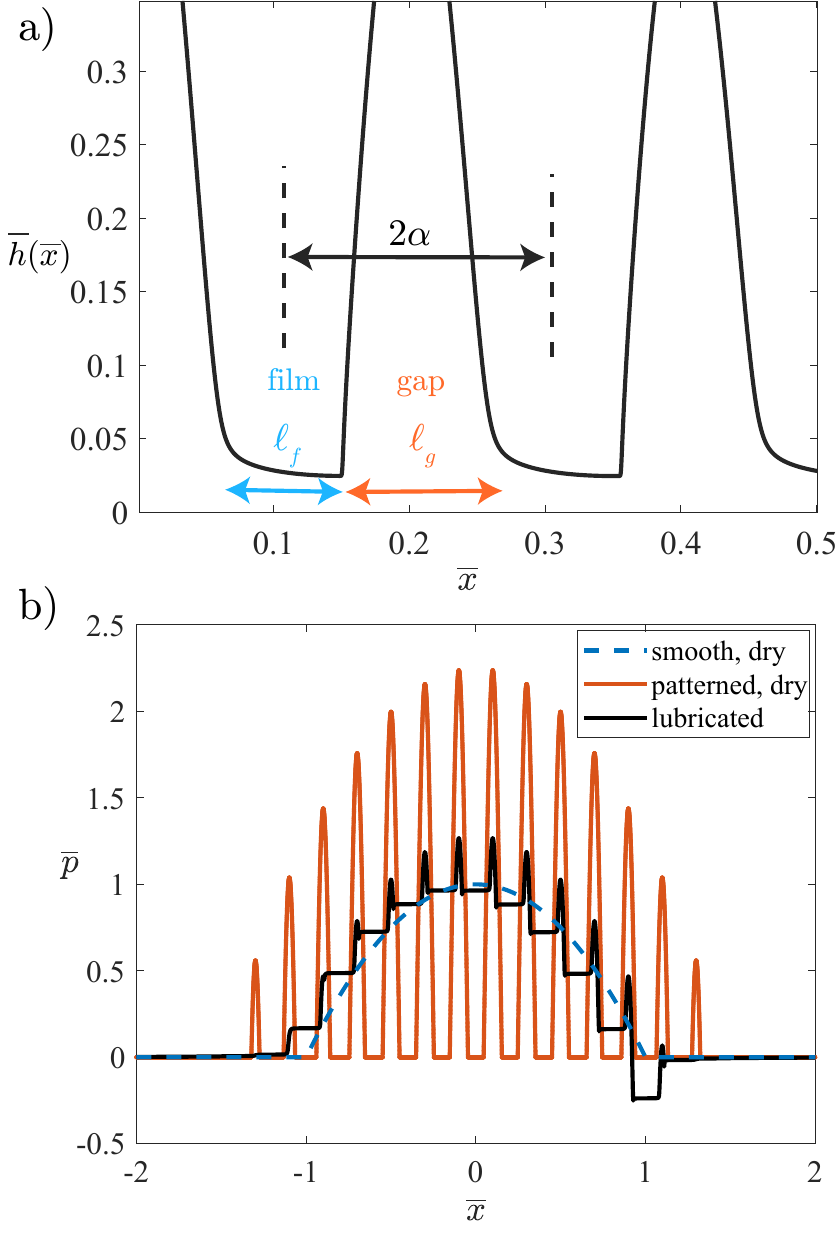}
    \caption{(a) A zoomed-in plot of the film thickness for $\beta=1.6$ and $\lambda=0.05$, highlighting the film and gap regions. (b) dimensionless pressure $\overline{p}$ versus dimensionless horizontal coordinate $\overline{x}$ for the $\beta = 4$ and $\lambda = 0.05$, the corresponding dry contact, and the dry smooth contact.}
    \label{fig.dry}
\end{figure}
We now use \eqref{lubrication_eq} along with the conclusion $d \overline{p}/d \overline{x} = O(1)$ to estimate the film thickness as $\overline{h} \sim \lambda^{1/2}$, valid in the thin film regions. Referring back to \eqref{friction}, it is easy to see that all three stress contributions in the integrand scale as $\lambda^{1/2}$ in the film regions. Recalling that the drag force is mostly accumulated in $n$ thin-film regions, each of width $\overline{\ell_f}$ (cf. Figure \ref{fig.cum_int}a), we conclude that the drag force scales as $\lambda^{1/2} n \overline{\ell_f}$. Referring back to Figure \ref{fig.dry}b, we see that the number of teeth in contact $n$ and the length of the flat regions  $\overline{\ell_f}$ in the lubricated problem are determined by the dry contact features. In the appendix, we show using scaling arguments that the dry contact is characterized by $n \sim \beta^{1/8}/\alpha$, and $\overline{\ell_f} \sim \alpha \beta^{-3/8}$. Combining these arguments, we obtain the scaling relation $D \sim \lambda^{1/2} \beta^{-1/4}$. The coefficient of friction is therefore
\begin{align} \label{LargeBetaFriction}
    \mu \sim \frac{\ell}{2R} \lambda^{1/2} \beta^{-1/4}, \quad \mbox{for} \quad \lambda \ll 1, \quad \beta \gg 1,
\end{align}
and is validated by numerical results in Figure \ref{fig. mu_lambda}. Notably, the prediction is also independent of $\alpha$, consistent with the numerical  findings shown in Fig. \ref{fig.mu_lambda_alpha}(a).

\begin{comment}
   \begin{figure}[t!]
\begin{subfigure}{0.4 \textwidth}
    \centering
    \includegraphics[width=0.95\textwidth]{fig_cum_D_f.pdf}
\end{subfigure}
\begin{subfigure}{0.4 \textwidth}
    \centering
    \includegraphics[width=0.95\textwidth]{fig_cum_L_f.pdf}
\end{subfigure}
    \caption{Plot of cumulative dimensionless forces: (a) drag force and (b) lift force in blue, together with film thickness $\overline{h}$  in red as a function of dimensionless horizontal coordinate ($\overline{x}$) for $\lambda = 0.05$ and $\beta = 4.0$. The plots feature two vertical axes: the left axis indicates the cumulative integral values, while the right axis displays the dimensionless film thickness $\overline{h}$}
    \label{fig.cum_int}
\end{figure}    
\end{comment}
\begin{figure}[t!]
    \centering
    \includegraphics[width=0.45\textwidth]{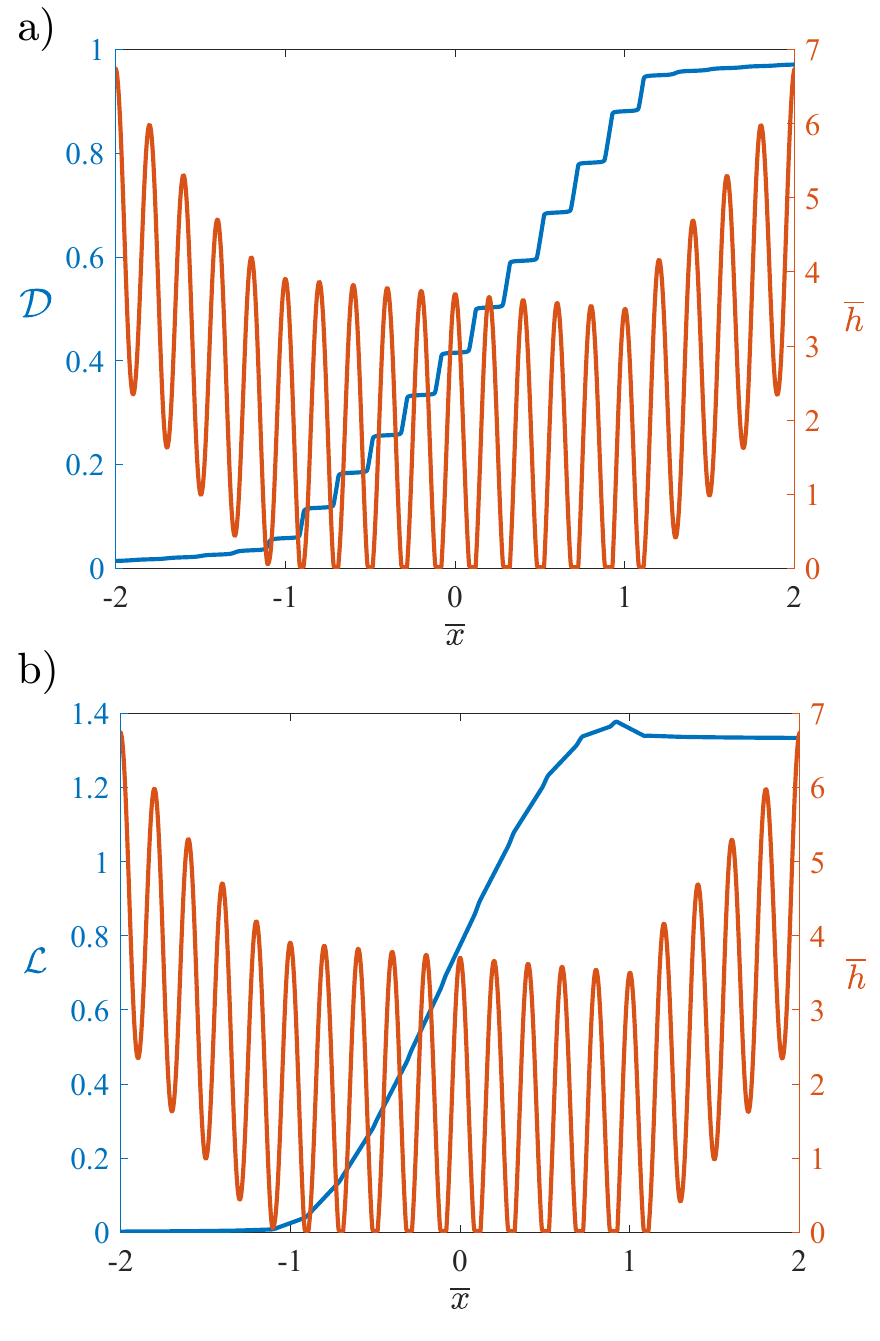}
    \caption{Plot of cumulative dimensionless forces: (a) drag force and (b) lift force in blue, together with film thickness $\overline{h}$  in red as a function of dimensionless horizontal coordinate ($\overline{x}$) for $\lambda = 0.05$ and $\beta = 4.0$. The plots feature two vertical axes: the left axis indicates the cumulative integral values, while the right axis displays the dimensionless film thickness $\overline{h}$}
    \label{fig.cum_int}
\end{figure}

%%%%%%%%%%%%%%%%
%%%%%%%%%%%%%%%%%%%%%%%%%
%%%Asymptotic Analysis Figure
\begin{comment}
 \begin{figure}[t!]
\begin{subfigure}{0.45\textwidth}
     \centering
      \includegraphics[width=0.95\linewidth]{fig_asymp_lambda.pdf}
      \label{fig.asymp_lambda}
      %\caption{}
\end{subfigure}
\begin{subfigure}{0.45\textwidth}
     \centering
      \includegraphics[width=0.95\linewidth]{fig_asymp_beta.pdf}
      \label{fig.asymp_beta}
      %\caption{}
\end{subfigure}
\caption{(a) Rescaled friction coefficient, $\mu (2R)/\ell$, versus the dimensionless speed, $\lambda$. Solid lines depict the numerical results, while dashed lines illustrate the analytical predictions. (b) The deviation of the friction coefficient from the soft smooth problem, $\mu - \mu_0$, presented against the dimensionless surface roughness, $\beta$, for a fixed dimensionless speed of $\lambda = 1000$. $\alpha =0.6$ throughout the figure.}
\label{fig.asymp}
\end{figure}   
\end{comment}

\begin{figure}[t!]
    \centering
    \includegraphics[width=0.45\textwidth]{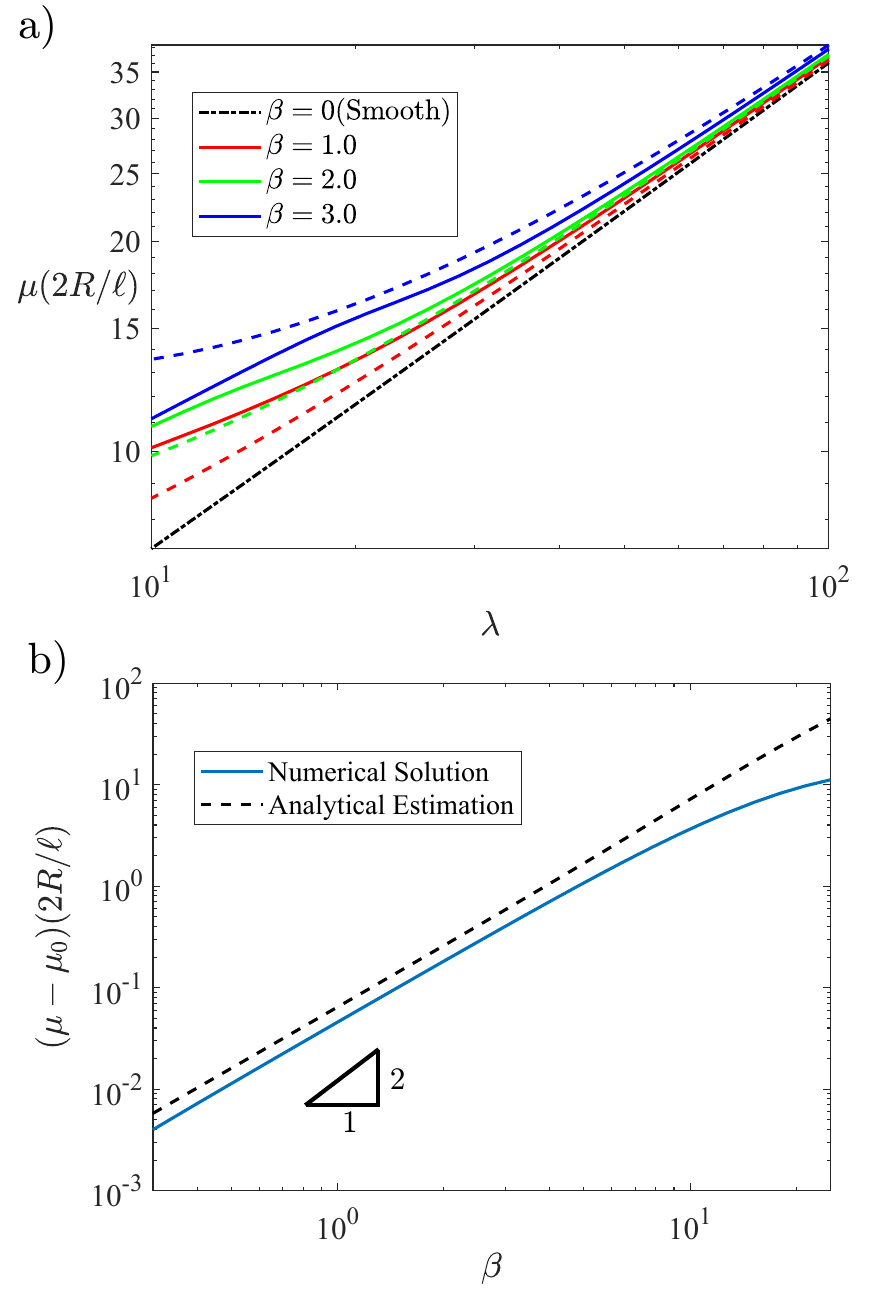}
    \caption{(a) Rescaled friction coefficient, $\mu (2R)/\ell$, versus the dimensionless speed, $\lambda$. Solid lines depict the numerical results, while dashed lines illustrate the analytical predictions. (b) The deviation of the friction coefficient from the soft smooth problem, $\mu - \mu_0$, presented against the dimensionless surface roughness, $\beta$, for a fixed dimensionless speed of $\lambda = 1000$. $\alpha =0.6$ throughout the figure.}
    \label{fig.asymp}
\end{figure}

%%%%%%%
%%%%%%%%%%%%

\subsection{Friction at high speeds: perturbation theory} \label{SecLargeLambda}
For large $\lambda$, the film thickness increases (Fig. \ref{fig.h and p}a), and eventually becomes much greater than both the elastic deformation and the roughness amplitude. We thus employ a small-deformation perturbation expansion in the vein of \citet{skotheim2004}, but include roughness effects. We outline the main ideas here, and present a more detailed analysis in appendix \ref{FixedIndentApp}.  For large $\lambda$, the rough cylinder floats above the undeformed surface of the bottom substrate, such that $c$ becomes large and negative. The ratio of the deformation scale to the film thickness is $\Lambda = \eta v_p K (2 R)^{1/2} |c|^{-5/2}$, which is small. Additionally, the roughness amplitude relative to the gap defines $\epsilon = b/(2|c|)$, which is also small. We expand the governing equations in a perturbation series for small $\Lambda$ (appendix \ref{FixedIndentApp}), solve the resulting problems numerically (for different values of $\epsilon$) and compute drag $D$ and lift $L$ forces. Separately, a small-roughness expansion of these coefficients identifies that the drag and lift take the approximate form $D = \eta v_p (2 R/|c|)^{1/2} (\mathfrak{D}_{0}  + \epsilon^2 \mathfrak{D}_{2})$ and $L = \Lambda \eta v_p (2 R/|c|) (\mathfrak{L}_{0}  + \epsilon^2 \mathfrak{L}_{2})$, up to corrections of $O(\Lambda^2, \epsilon^4)$ where the $\mathfrak{D}_{i}$ and $\mathfrak{L}_i$ are $O(1)$ parameter-independent coefficients (appendix \ref{FixedIndentApp}). 
%As noted earlier, in the limit where roughness amplitude is much smaller that fluid film thickness ($\beta \ll \overline{h}$) i.e large $\lambda$ in Fig. \ref{fig.mu_lambda_alpha}b, perturbation expansion can be employed to find an analytical solution for friction coefficient ($\mu$). 
%Comparing the numerical computation and the general analytic form identifies the functions $\mathfrak{L}(\epsilon)$ and $\mathfrak{D}(\epsilon)$ numerically. To gain analytic insight into the structure of these functions for small $\epsilon$, we expand them as  $\mathfrak{D} \sim   \mathfrak{D}_{0}  + \epsilon^2 \mathfrak{D}_{2}$,  $\mathfrak{L} \sim   \mathfrak{L}_{0}  + \epsilon^2 \mathfrak{L}_{2}$. 
The coefficients $\mathfrak{D}_0 = 2 \pi$ and $\mathfrak{L}_0= 3 \pi /8$ correspond to drag \citep{happel2012} and lift \citep{skotheim2004} on a smooth cylinder near a weakly deformable soft surface and are known analytically. The absence of terms linear in $\epsilon$ (and indeed all odd powers of $\epsilon$) stems from the symmetry of the problem, as noted in Sec. \ref{SecSmallLam}. We then find $\mathfrak{D}_2\approx 3 \pi /2$ and $\mathfrak{L}_2 \approx 3 \pi /2$ by fitting the analytic forms for $D$ and $L$ to our numerical force calculations. 

Having obtained forces in terms of the gap height $|c|$, we recast the expression for lift  in terms of the original ``fixed-load'' dimensionless variables to obtain 
\begin{equation}\label{algebraic_c}
    \frac{1}{48} \frac{\lambda^2}{|\overline{c}|^{7/2}} \left(\mathfrak{L}_0 + \frac{\beta^2}{|\overline{c}|^2} \mathfrak{L}_2\right)\,=1,
\end{equation}
which is a nonlinear equation for $|\overline{c}|$. Substituting the small-$\Lambda$, small-$\epsilon$ expressions for $D$ and $L$, we find after some manipulation that the coefficient of friction is 
%Next, we transform back \eqref{mu_asymp} to the constant-load system to compare our analytical solution with \eqref{friction}b,
\begin{align}\label{friction_norm_asymp}
    \mu &= \frac{\ell}{2R}  \frac{\lambda \left(\mathfrak{D}_0 + \mathfrak{D}_2 \frac{\beta^2}{\overline{c}^2}\right)}{8\, \overline{c}^{1/2}} .
\end{align}
Thus, we first solve the nonlinear equation \eqref{algebraic_c} for $\overline{c}$ and then substitute these values into \eqref{friction_norm_asymp} to obtain a semi-analytic approximation for the coefficient of friction $\mu$. A comparison of this result with the fully numerical calculation is depicted in Fig. \ref{fig.asymp} for different values of $\lambda$ and $\beta$. As evident from Fig. \ref{fig.asymp}(a), the results of  \eqref{friction_norm_asymp}, depicted by dashed lines, are in quantitative agreement with the numerical results at large $\lambda$. Moreover, the semi-analytic solution provides a reasonably accurate qualitative prediction at moderate speeds ($\lambda \approx 10$). 

Furthermore, it is clear from \eqref{algebraic_c} and \eqref{friction_norm_asymp} that the leading effects of roughness on $\mu$ scale with $\beta^2$. To examine the  effect of roughness on the friction coefficient, we subtracted the friction coefficient of the smooth problem, $\mu_0$, from the total friction coefficient $\mu$, and compared the higher-order terms in Fig. \ref{fig.asymp}(b) at fixed $\lambda$. The $\beta ^2$ scaling (dashed curve) is evident in the figure, with analytical results showing good agreement with numerics up to $\beta \approx 10$.

%As previously noted, $D_{0}$ and $L_{0}$ in \eqref{friction_norm_asymp} are derived from the well established results in previous studies and represent the soft smooth condition where the top surface in Fig. \ref{Schematic} does not have any protrusions/teeth ($b=0$), in contrast, $D_2$ and $L_2$ are introduced for the first time in this work. 

%%%%%%%%%%%%%%%%%%
%%%%%%%%%%
\section{Conclusions}
We have studied the lubricated contact of a patterned curved surface on a soft coated substrate, assuming a local Winkler-type elastic description. The model is governed by three dimensionless parameters: $\alpha$, the dimensionless wavelength of the patterns; $\beta$, the dimensionless roughness amplitude; and $\lambda$, the dimensionless relative speed of the contacting surfaces. Numerical solutions and symmetry arguments show that the effective (lubricated) coefficient of friction is insensitive to variations in $\alpha$ for small $\alpha$. %We found that the coefficient of friction increases with speed, similar to the case of smooth surfaces. 

An interesting and nontrivial finding is that the coefficient of friction depends non-monotonically on the roughness amplitude at fixed speeds. For small roughness amplitudes, the friction coefficient grows from the smooth case proportional to $\beta^2$, which we understand using symmetry arguments. For large roughness, the friction coefficient decreases as $\beta^{-1/4}$, eventually becoming lower than even the smooth limit. We understand this behavior for small speeds by analyzing the contact geometry established in the ``dry'' static limit. While much of the drag is accounted for by thin films, the applied load is counteracted by the spaces in between the films. At larger roughness, the fraction of contact occupied by the films decreases, leading to lower drag. At much larger speeds, a greater volume of fluid is entrained leading to thicker films, and the effects of surface roughness become weaker.   

The theoretical and numerical insights provided by this work reveal interesting and non-intuitive behaviors arising from interactions between fluid flow, elasticity and the multiple length scales associated with patterned surfaces. Among these is the non-monotonic dependence of the friction coefficient on the amplitude of the pattern. This behavior is reminiscent of (albeit distinct from) the recent experimental observations of \citep{peng2021}, where a non-monotonic dependence of friction on sliding speed was observed for patterned surfaces, but was absent for smooth surfaces. The insights identified by this work offer a deeper understanding of the role of surface patterns and roughness in the interactions of soft materials. We envision further development of our findings to more complex geometric or material configurations, as well their use in the design of haptic and soft robotic applications.

\section*{Author contributions}
A. K.-E. developed the theory and simulations. B. R. developed scaling arguments for dry contact. Both authors conceived the project, analyzed the results and wrote the paper. %We strongly encourage authors to include author contributions and recommend using \href{https://casrai.org/credit/}{CRediT} for standardised contribution descriptions. Please refer to our general \href{https://www.rsc.org/journals-books-databases/journal-authors-reviewers/author-responsibilities/}{author guidelines} for more information about authorship.

\section*{Conflicts of interest}
The authors have no conflicts of interest. 
%In accordance with our policy on \href{https://www.rsc.org/journals-books-databases/journal-authors-reviewers/author-responsibilities/#code-of-conduct}{Conflicts of interest} please ensure that a conflicts of interest statement is included in your manuscript here.  Please note that this statement is required for all submitted manuscripts.  If no conflicts exist, please state that ``There are no conflicts to declare''.

\begin{comment}
\section*{Data availability}

A data availability statement (DAS) is required to be submitted alongside all articles. Please read our \href{https://www.rsc.org/journals-books-databases/author-and-reviewer-hub/authors-information/prepare-and-format/data-sharing/#dataavailabilitystatements}{full guidance on data availability statements} for more details and examples of suitable statements you can use.    
\end{comment}

\section*{Acknowledgements}
The authors acknowledge the National Science Foundation for support through awards CBET-2328628. A. K.-E. thanks the US Department of Education for support through a GAANN Fellowship (P200A210080). 
%%%END OF MAIN TEXT%%%

%%%%%
\begin{appendix}

\section{Dry contact at large $\beta$}
As observed in Figure \ref{fig.dry}b, when $\lambda \ll 1$, the lubricated problem inherits certain features from the dry problem, such as the number of teeth in contact $n$, the length of flattened regions $\ell_f$, and the gap regions $\ell_g$. Here, we focus on the limit where $\beta$ is large and study the scaling of various quantities with respect to  $\alpha$ and $\beta$. 

At dry contact, the teeth ``penetrate'' into the soft material a depth $c$. This sets up two horizontal length scales: a global ``apparent contact'' length scale $(R c)^{1/2}$, and a flattened local contact scale $\ell_f \sim (r_f c)^{1/2}$ on each tooth, where $r_f$ is the radius of curvature of the tooth. By taking the second derivative of \eqref{Geometry}, we find that $r_f \sim a^2 /b$, so the dimensionless local contact length is $\overline{\ell}_f = \ell_f/\ell \sim \alpha (\overline{c}/\beta)^{1/2}$. The global length scale is spread across $n$ teeth a distance $a$ apart from each other, so $(R c)^{1/2} \sim a n$, or in dimensionless terms $\overline{c}^{1/2} \sim \alpha n$. Additionally, using \eqref{lift_norm} and considering that the load is supported by the flattened contact regions, we can express $\overline{p} \overline{\ell}_f n  \sim 1$. For a Winkler-elastic solid, $\overline{p} \sim \overline{c}$. Combining these scaling relations, the geometric features of dry contact scale with $\alpha$ and $\beta$ as
\begin{equation}\label{dry_scaling}
\overline{c} \sim \beta^{1/4}, \qquad \overline{\ell}_f \sim \alpha \beta^{-3/8}, \qquad n \sim \alpha^{-1 }\beta^{1/8}
\end{equation}
Notably, the product $n \overline{\ell}_f \sim \beta^{1/4}$ (independent of $\alpha$), which enters \eqref{LargeBetaFriction}. To validate this scaling analysis, Figure \ref{fig.dry_n_lf} compares $n \overline{\ell}_f$, the variable employed in the main text, from \eqref{dry_scaling} with the numerical solution of the dry contact problem, obtained from solving \eqref{Geom_norm} with $\overline{h}=0$, subject to \eqref{lift_norm}.
\begin{figure}
    \centering
    \includegraphics[width=0.45\textwidth]{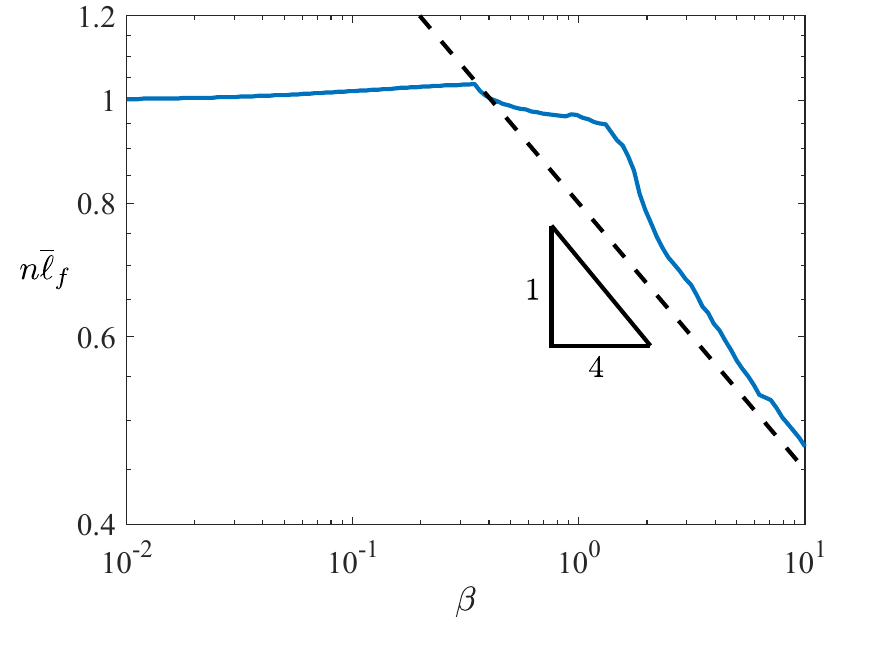}
    \caption{Plot of $n \ell_f$ versus $\beta$ for the dry contact with $\alpha = 0.1$. The solid line represents the numerical solution, while the dashed line illustrates the scaling prediction.}
    \label{fig.dry_n_lf}
\end{figure}
\section{Large-$\lambda$ perturbation expansion} \label{FixedIndentApp}
At large $\lambda$, the film thickness is greater than both the deformation and the roughness amplitude, making the problem amenable to a perturbation analysis. Extensively discussed for smooth surfaces in existing literature \citep{skotheim2004,kargar2022}, we adapt this approach to patterned surfaces.   It is convenient to reformulate the problem in terms of the separation $|c|$ rather than the load, and later translate back to the fixed-load framework introduced in Sec. \ref{SecFormulation}. The film thickness scales with $|c|$, setting the horizontal length scale $\sqrt{2R|c|}$. The relevant pressure scale is $\eta v_p R^{1/2} |c|^{-3/2}$, %Note that in this formulation, the separation is held fixed, but the normal load may vary. 
which, through \eqref{elastic_response}, identifies the deformation scale. Normalizing variables according to these scales leads to the rescaled lubrication equation (uppercase letters are dimensionless)
\begin{equation}\label{lubrication_mah}
\frac{\partial}{\partial X} \left( H^3 \frac{\partial P}{\partial X} - 6 H\right)=0.
\end{equation}
The fluid film thickness $H$ is expressed in terms of pressure and geometric parameters as:
\begin{equation}\label{film_thickness_mah}
    H(X)= 1+X^2+\epsilon \cos{\left(\frac{\pi X}{\hat{\alpha}}\right)} + \Lambda P(X),
\end{equation}
where $\Lambda$ is the dimensionless deformation, $\epsilon$ is dimensionless roughness amplitude (both defined in Sec. \ref{SecLargeLambda}), and $\hat{\alpha} = a/\sqrt{2 |c| R}$. All three parameters are small in the asymptotic sense.  

We develop a perturbation expansion for $\Lambda \ll 1$, expanding pressure as $P = P_0 + \Lambda P_1 + O(\Lambda^2)$. Substituting into \eqref{film_thickness_mah} yields 
\begin{subequations}
\begin{align}%\label{lubrication_eta_0}
   \Lambda^0:&\; \frac{\partial}{\partial X} \left( H_0^3 \frac{\partial P_0(X)}{\partial X} - 6 H\right)=0, \label{lubrication_eta_0}  \\
    \Lambda^1:&\; \frac{\partial}{\partial X} \left( H_0^3 \frac{\partial P_1(X)}{\partial X} + 3 H_0 P_0(X) \frac{\partial P_0(X)}{\partial X} - 6 P_0(x)\right)=0, \label{lubrication_eta_1}
\end{align}
\end{subequations}
 where $H_0 = 1+X^2+\epsilon \cos{\pi X/ \hat{\alpha}}$ represents the film thickness of the corresponding rigid problem ($\Lambda \to 0$).  We solve \eqref{lubrication_eta_0} and \eqref{lubrication_eta_1} numerically for different $\epsilon$. We use these solutions to compute  drag and lift forces, which (here written as dimensional quantities) %For the smooth problem the  drag scales as $\eta v_p (2 R/|c|)^{1/2}$ and the lift scales as $\eta v_p (2 R/|c|) \Lambda$, up to terms of $O(\Lambda^2)$ or smaller \cite{skotheim2004, rallabandi2017rotation}.
 take the form $D = \eta v_p (2 R/|c|)^{1/2} \mathfrak{D}(\epsilon)$ and $L = \Lambda \eta v_p (2 R/|c|) \mathfrak{L}(\epsilon)$, where $\mathfrak{D}(\epsilon)$ and $\mathfrak{L}(\epsilon)$ are $O(1)$ functions that result from the numerical computation. Noting that $\epsilon \ll 1$ for large $\lambda$ we expand these functions as  $\mathfrak{D}=\mathfrak{D}_0 + \epsilon^2 \mathfrak{D}_2 + \dots $ and  $\mathfrak{L}=\eta(\mathfrak{L}_0 + \epsilon^2 \mathfrak{L}_2)$. The coefficients $\mathfrak{D}_0 = 2 \pi$ and $\mathfrak{L}_0= 3 \pi /8$ are well-established results in the literature \citep{happel2012,skotheim2004} and correspond to the smooth problem. A fit to the numerical force computations determine $\mathfrak{D}_2\approx 3 \pi /2$ and $\mathfrak{L}_2\approx 3 \pi /2$.

\end{appendix}


%%%REFERENCES%%%
% \bibliography{softroughlubrication} %You need to replace "rsc" on this line with the name of your .bib file
% \bibliographystyle{rsc} %the RSC's .bst file


\end{document}